\documentstyle[12pt,aaspp4,psfig]{article}
\slugcomment{To Appear in ApJ Letters (accepted Jul. 24, 1997)}
\def\AL {$\alpha $}                    
\def\MOLH {\hbox{${\rm H}_2$}}         
\def\THCO {\hbox{$^{13}{\rm CO}$}}     
\def\CNOA {\hbox{C91$\alpha$}}         
\def\CSIA {\hbox{C65$\alpha$}}         
\def\CFINE {\hbox{[C {\sc ii}] 158~$\mu$m}}    
\def\kms{km s$^{-1}$}
\def\mic {$\mu\hbox{m}$}
\def\percc {$\hbox{{\rm cm}}^{-3}$}    
\def\THEC  {$\Theta _1$C Ori}          
\def\ILIN  {\hbox{T$_{l}\, \Delta $v}} 
\begin{document}

\title{Carbon radio recombination lines in the Orion Bar}

\author{F.~Wyrowski\altaffilmark{1}, P.~Schilke\altaffilmark{1,2},
        P.~Hofner\altaffilmark{3}, and C.M.~Walmsley\altaffilmark{4}}

\altaffiltext{1}{I. Physikalisches Institut, Univ. zu K\"{o}ln, 
                 Z\"{u}lpicherstr. 77, D-50937 K\"{o}ln, Germany}
\altaffiltext{2}{Max-Planck-Institut f\"ur Radioastronomie,
                               Auf dem H\"ugel 69, D-53121 Bonn, Germany}
\altaffiltext{3}{Cornell University, NAIC, Arecibo Observatory, P.O. Box
           995, Arecibo, P.R. 00613, USA}
\altaffiltext{4}{Osservatorio Astrofisico di Arcetri, Largo E.Fermi 5,
        I-50125 Firenze, Italy}

%
%
\begin{abstract}
  We have carried out VLA D-array observations of the \CNOA\ carbon
  recombination line as well as Effelsberg 100-m observations of the
  \CSIA\ line in a 5\arcmin\ square region centered between the Bar and
  the Trapezium stars in the Orion Nebula with spatial resolutions of
  10\arcsec\ and 40\arcsec, respectively.  The results show the
  ionized carbon in the PDR associated with the Orion Bar to be in a
  thin, clumpy layer sandwiched between the ionization front and the
  molecular gas. From the observed line widths we get an upper limit
  on the temperature in the C$^{+}$ layer of 1500~K and from the
  line intensity a hydrogen density between $5\times 10^4$ and
  $2.5\times 10^5$~\percc\ for a homogeneous medium.  The observed
  carbon level population is not consistent with predictions of
  hydrogenic recombination theory but could be explained by
  dielectronic recombination. The layer of ionized carbon seen in
  \CNOA\ is found to be essentially coincident with emission in the
  v=1-0 S(1) line of vibrationally excited molecular hydrogen. This is
  surprising in the light of current PDR models and some possible
  explanations of the discrepancy are discussed.

\end{abstract}

%
%
  \section{Introduction}
  
  The neutral gas surrounding \ion{H}{2} regions is an excellent
  laboratory for the purpose of understanding the interactions between
  the UV radiation of young newly formed stars and the surrounding
  molecular clouds.  Such photon dominated regions (PDR's) have been
  the subject of both theoretical (e.g.\ Tielens \& Hollenbach 1985,
  Sternberg \& Dalgarno 1995) and observational attention (e.g.\ 
  Herrmann et al. 1996).  The most easily observable example of a PDR
  is perhaps that provided by the Orion Nebula and the surrounding hot
  neutral gas.  Of particular interest is the elongated bar--like
  feature to the south--east of the Orion Trapezium stars known as the
  Orion Bar (see Tielens et al.\ 1993, Hogerheijde et al.\ 1995, van
  der Werf et al.\ 1996 for recent discussions).  The Bar is elongated
  in a direction on the plane of the sky almost perpendicular to the
  line of sight to the O6 star $\Theta ^{1}$C Ori in the trapezium,
  which is responsible for ionizing the \ion{H}{2} region and heating
  the neutral gas.  This fact makes it possible for us to observe
  stratification within the Bar along the line of sight to $\Theta
  ^{1}$C Ori.  Thus, the Bar seen in free--free emission from the
  ionized gas is clearly offset with respect to the Bar seen in
  rovibrational lines of \MOLH\ and fine structure lines of \ion{C}{2}
  and \ion{O}{1} from the neutral gas.  These in turn are offset
  relative to the Bar as seen in various molecular species (see
  Tielens et al.\ 1993, Herrmann et al.\ 1996 for examples). It has
  been concluded that the observed stratification is consistent with a
  mean hydrogen density of $5\times 10^4$ \percc.  On the other hand,
  high angular resolution maps of the Bar in various molecular
  transitions (Tauber et al.\ 1994) have shown that the neutral gas is
  structured and quite inhomogeneous with clump densities as high as
  $10^6$ \percc .
  
  The carbon radio recombination lines offer an alternative approach
  to studying PDR's. Their emission is invariably optically thin and
  proportional to the \emph{square} of the electron density (or
  equivalently of the carbon abundance in the neutral regions where
  carbon is singly ionized).  Natta et al.\ (1994) demonstrate
  moreover that one can use the ratio of the radio line intensity to
  that of the far infrared \CFINE\ line to infer the density in the
  layer where carbon is singly ionized.  They observed \CNOA \ with
  the Effelsberg 100-m telescope at several positions in Orion and
  derived densities of $10^6$~\percc.  These observations however were
  made with a relatively poor angular resolution (80\arcsec ) which
  rendered difficult the comparison with other PDR tracers in Orion.
  We have therefore undertaken new carbon radio line observations with
  the objective of allowing a comparison both with far infrared and
  molecular line data.  In the first place, we have mapped the \CSIA \ 
  line using the Effelsberg 100-m telescope at an angular resolution
  of 40\arcsec .  This allows a reasonable comparison with the far
  infrared fine structure line maps of Stacey et al. (1993).
  Secondly, we have used the VLA to map \CNOA\ towards the Bar with
  10\arcsec \ resolution.  This can usefully be compared with the
  H$_2^*$ maps of van der Werf et al.\ (1996) as well as with recent
  studies of \THCO (3--2) made with the CSO telescope (Lis et al.
  1997).  In this paper, we present the results of the carbon radio
  line mapping and show that the stratification seen in other species
  is also clearly present in the VLA \CNOA \ maps.

%
  \section{Observations}

    \subsection{\CSIA\ observations of Orion with
      the Effelsberg 100-m telescope}
           
    The observations were carried out using the Effelsberg 100-m
    telescope on 4 sessions in September 1993, March 1994, January and
    December 1996.  Pointing was tested on 3C120 and 3C161 at hourly
    intervals and has an accuracy of 5\arcsec . The facility K-band
    maser was used with system temperatures of 150 to 200~K.  We
    observed a 5\arcmin\ square region centered between the Bar and
    the Trapezium in a total power mode with reference positions
    75\arcmin\ to the east.  The spectrometer was a autocorrelator
    which was split into two sections of 512~channels and centered on
    \CSIA\ (23415.9609~MHz) and H65\AL\ (23404.2793~MHz).  Each
    section had a bandwidth of 12.5~MHz yielding a spectral resolution
    of 0.3~\kms. The telescope half-power beamwidth at 23.416~GHz is
    40\arcsec.

    \subsection{\CNOA \ observations of the Orion Bar with
      the VLA}
    
    We observed the Orion Bar in July 1996 using the VLA in its D
    configuration.  The basic observing parameters were as given in
    Table 1. The phase center of our map was chosen to be slightly
    offset from the Bar in order to map also the total continuum
    radiation of the Orion~A \ion{H}{2} region.  We observed a
    bandpass calibrator at intervals of 3~hours and made phase
    calibrations every 25~minutes. The data were processed using AIPS
    and resulted in synthesized beams of
    11.7\arcsec$\times$9.0\arcsec\ with natural weighting and a final
    RMS noise of 2~mJy per beam (0.3~K) in a channel map.  Our main
    difficulty is separating the blend of He91$\alpha $ \ and \CNOA\ 
    which is done by subtraction of a linear spectral baseline from
    every data point in the uv plane using small windows in the
    immediate neighborhood of the carbon line.  This works well for
    the Bar but could lead to overestimation of the line emission
    where the helium line dominates the spectra. A 3.5~cm continuum
    image was produced by summing over the line free channels.
    Comparing \CNOA\ VLA observations smoothed to the corresponding
    100-m beam (80\arcsec) with \CNOA\ 100-m observations carried out
    in August 1996 of the Bar, we found that we underestimate the
    \CNOA\ intensity due to the missing short spacing information by
    less than 30\%.
 
  \section{Observational Results}
  
    \subsection{Morphology of carbon radio line emission}
 
    In Fig.~1, our \CNOA\ (VLA) and \CSIA\ (100-m) recombination line
    maps are shown compared with each other.  Despite the difference
    in angular resolutions, one notes that the same basic features are
    seen on both maps.  One sees the Bar to the south-east, a compact
    (1 arc minute in size) region 30\arcsec \ to the south--west of
    \THEC \ in the central portion of the map, and another feature
    close to the BN-KL region on the northern edge of the map. 
  
    It is interesting to compare our \CSIA\ (100-m) integrated
    intensity map with the \CFINE \ contours from Stacey et al.
    (1993).  In this case, both lines are probing ionized carbon and
    with similar angular resolution. It is striking to note again the
    similarity of the general characteristics of these maps in spite
    of different excitation characteristics (see Natta et al.  1994
    ). Thus, we conclude that the FIR and radio lines come
    from essentially the same regions in space although the
    recombination line shows a higher contrast and there are likely to
    be differences on scales of a few arc seconds.
 
    That the regions seen in the carbon radio lines represent PDR's in
    the vicinity of ionization fronts is demonstrated by Fig. 2.  It
    shows our VLA \CNOA \ map compared with our 3.5 cm continuum map
    representing the distribution of free--free emission from the
    ionized gas.  The Bar as seen in \CNOA \ is parallel to but offset
    by about 20\arcsec \ (0.05 pc) from the Bar of ionized gas.  This
    is in the direction away from \THEC \ and is thus consistent with
    ionization and heating from that source. The distribution of \THCO
    (3--2) observed by Lis et al.  (1997) using the CSO with a
    20\arcsec \ HPBW is offset a further 20\arcsec\ to the SE relative
    to \CNOA \ and thus, the carbon radio line emission is neatly
    sandwiched between the ionized and molecular media.
    
    A sample spectrum of the \CNOA\ line towards the Bar is shown as
    an insert in Fig.~2 compared with the corresponding \THCO\ 
    spectrum.  One sees from this that the line profiles are very
    similar indicating a close link between the two. This kinematical
    agreement between molecular gas and carbon emission is found
    nearly everywhere showing the \ion{C}{2} region to be very
    quiescent with only small velocity shifts.  At most positions, the
    line emission is narrow with line widths between 2 and 2.5 \kms.
    The VLA linewidths can be converted into upper limits on the
    kinetic temperature between 1000 and 1600~K in the region
    responsible for the \CNOA \ emission in the Bar (assuming thermal
    broadening to be solely responsible for the observed line width).
    The carbon lines are broader towards the clump of \CNOA \ emission
    to the SW of the Trapezium where we find an upper limit to the
    temperature of 3000 K.
  
    \subsection{Level populations in the \CNOA \  emitting region }
 
    By smoothing our VLA \CNOA \ data to 40\arcsec \ resolution, we
    can make a direct estimate of the ratio of the intensities of the
    two recombination lines at several positions and Table 2 gives the
    results. We also give the \CFINE \ intensities at the same
    positions derived from the maps of Stacey et al. (1993).  
    
    Table 2 shows that we observe typical line intensity ratios \ILIN
    (\CNOA )/ \ILIN (\CSIA ) between 2.3 and 3.3 and ratios \ILIN
    (\CNOA )/I(\CFINE ) of 800 K \kms~erg$^{-1}$~cm$^2$~s~sr.  One can
    compare the former with the ratios expected for hydrogenic
    departure coefficients ($b_{n}$'s) which, for example, are 5 for
    T=800~K and an electron density n$_{e}$ of 10~\percc \ (see Natta
    et al.  Fig.~7). Given our limits on the electron temperature and
    for reasonable electron densities, it is not possible to get
    agreement between observations and the line ratios predicted by
    hydrogenic recombination theory.  We believe that the explanation
    of this is that dielectronic recombination of the type discussed
    by Walmsley \& Watson (1982) causes the populations to be closer
    to being thermalised than hydrogenic theory predicts. If one goes
    to the extreme of thermalised level populations ($b_{n}=1$), the
    predicted \ILIN (\CNOA )/ \ILIN (\CSIA ) is 2.7.  A more
    reasonable assumption might be that atoms with a $^{2}P_{3/2}$
    core have thermalised populations whereas atoms with $^{2}P_{1/2}$
    cores are hydrogenic. Then, applying Eq.~8 from Walmsley \&
    Watson(1982), we find \ILIN (\CNOA )/ \ILIN(\CSIA ) is 3.1 for
    $T$=200~K and $n=10^5$~\percc. Given the errors, both of these are
    consistent with the observations and we conclude that the
    dielectronic process drives the level populations close to the
    values expected in LTE. In the following discussion, we have for
    simplicity set the $b_{n}$'s equal to unity .

    \subsection{Comparison with molecular hydrogen emission} 
    
    Hot neutral gas in the neighborhood of ionization fronts is also
    traced by emission in the NIR 2\mic \ lines of \MOLH \ which are
    collisionally excited at temperatures upwards of 1500~K
    although one may observe fluorescence at low temperatures.  
    It is thus of interest to compare our VLA map of \CNOA \ with the
    emission in the 1-0 S(1) line of \MOLH \ which was imaged by van
    der Werf et al.  (1996).  The qualitative result of this is that
    the Bar as seen in \MOLH \ corresponds generally with the carbon
    radio lines when one takes into account the difference in angular
    resolutions.
 
    A closer view reveals that the carbon radio line emission appears
    to peak slightly closer to \THEC \ than does the emission from
    vibrationally excited molecular hydrogen.  One obtains some
    insight into the situation from Fig.~3 which shows the results of
    an homogeneous edge-on model derived using the PDR models of
    Tielens \& Hollenbach (1985) with $n$(H)=$10^5$~\percc\ and
    $G_0=10^5$ extrapolated to a density of $7\times 10^4$~\percc.  We
    assume a perfectly edge-on model without a tilt.  The intensity of
    the \CNOA \ line, which is proportional to $T^{-1.5}$, increases
    with depth into the PDR and reaches a peak just in front of the
    transition zone C$^+$/\ion{C}{1}/CO.  The observed cross cuts of
    \CNOA \ and \MOLH \ v=1-0 S(1) shown in Fig.~3 are averaged over
    the Bar and the half-power point of the 3.5~cm continuum is used
    to define the ionization front. In contrast to the prediction of
    the homogeneous model, the observed \CNOA\ emission is broader and
    extends farther into the Bar.  But while the predictions for \CNOA
    \ are in qualitative agreement with observation, the same is not
    true for molecular hydrogen which is expected by theory to lie
    between ionization front and \CNOA.
 
    The explanation of this is unclear but one possibility is that the
    heat input to the observed \MOLH \ emission is due to a shock.
    One difficulty with this interpretation (see Hill and Hollenbach
    1978, Tielens et al. 1993) is the general quiescence of the gas in
    the Bar discussed in Sect.~3.1.  However, explaining our result
    seems to require either considerable heat input or fluorescence in
    a layer at a depth in the Bar corresponding to roughly 5
    magnitudes of visual extinction.

    \subsection{Physical parameters in the \CNOA \ emitting region}
    
    We now summarize briefly the constraints on density and
    temperature in the layer of the Bar emitting \CNOA . As noted in
    Sect.~3.1, we can place a limit of 1600~K on the kinetic
    temperature.  In order to explain the observed \CNOA \ line
    intensity (5~K~\kms\ peak value), we find (taking $b_{n}=1$, see
    last section) that the carbon line emission measure $E_{l}=\int
    n_{e}\, n(C^{+})\, ds$ (in pc~cm$^{-6}$) is equal to $2250\,
    T_{3}^{1.5}$ where $T_{3}=T/1000$.  For an Orion carbon abundance
    [C]/[H] = $3.4\times 10^{-4}$ (Peimbert 1993) and assuming $n_{e}
    = n({\rm C}^{+})$, we estimate the hydrogen density in the layer
    responsible for the observed \CNOA \ emission to be $n_{\rm H}=
    4\cdot 10^5\, T_{3}^{.75}\, L_{\rm B}^{-0.5}$ where $L_{\rm B}$
    (pc) is the depth of the Bar along the line of sight (of order
    0.6~pc according to Hogerheijde et al. 1995). Thus for
    temperatures between 200 and 1600~K and $L_{\rm B}= 0.6$, we find
    $n_{\rm H}$ between $5\times 10^4$ and $2.5\times 10^5$~\percc\ 
    assuming a beam filling factor of 1.  This is consistent with
    recent estimates of the density on basis of CN and CS observations
    with comparable resolution (Simon et al.  1997) and also
    consistent with our results of the previous section for a typical
    temperature in the \CNOA\ emission region of 300~K.
 
  \section{Conclusions}
   
  Our data show convincingly that carbon is ionized in a layer
  intermediate between molecular gas and ionization front. We put an
  upper limit on the temperature of the \ion{C}{2} layer of 1600~K and
  find that the hydrogen density must be between $5\times 10^4$ and
  $2.5\times 10^5$~\percc. We also have found evidence that the
  dielectronic recombination process discussed by Walmsley \& Watson
  (1982) plays an important role in populating high $n$ levels under
  the conditions of the Orion Bar PDR. Finally, we show that current
  models have difficulty in explaining the morphology of the
  vibrationally excited molecular hydrogen emission 
  relative to ionized carbon in the Orion Bar.

  \acknowledgments
  
  We acknowledge use of the VLA of the National Radio Astronomical
  Observatory which is operated by Associated Universities Inc. under
  cooperative agreement with the National Science Foundation.  Dr
  G.~Stacey kindly made available to us his \CFINE\ map of Orion.
  We thank the referee (D. Hollenbach) for his comments on the
  manuscript. C.M.W acknowledges the support from ASI grants 94-RS-152
  and ARS-96-66 and CNR grant 96/00317 for star formation research at
  Arcetri.


\begin{figure}[p]
  \hbox to \textwidth{
  \hfill  
    \vbox{
    \psfig{figure=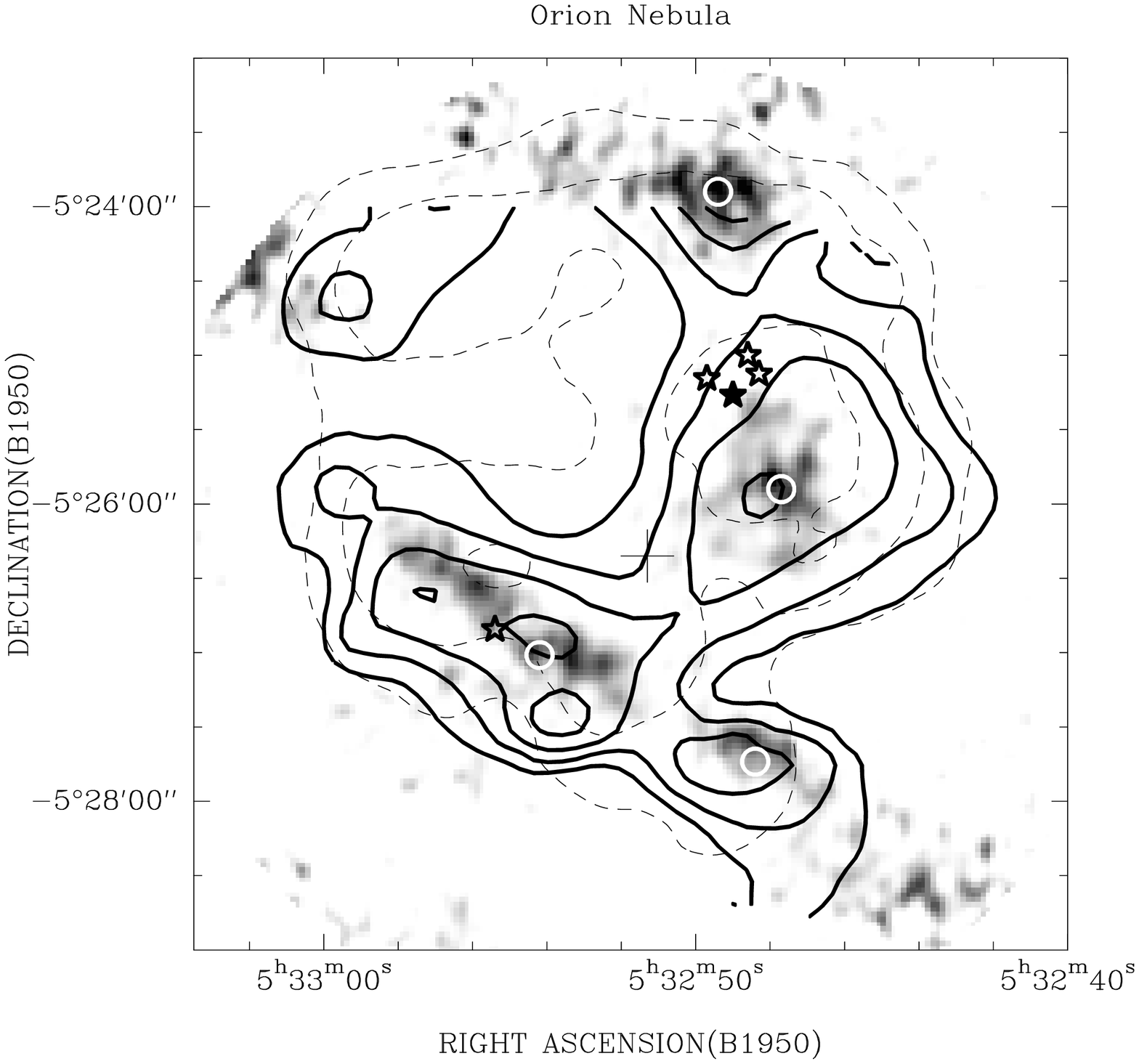,bbllx=35pt,bblly=10pt,bburx=560pt,bbury=670pt,width=18cm,angle=-90} 
         }
  \hfill
  }
    \caption
    {Comparison of \CNOA\ (VLA) integrated intensity (HPBW 10\arcsec,
      greyscale) and \CSIA\ (100-m) integrated intensity (40\arcsec \ 
      HPBW, bold contours, contours are 30, 50, 70, 90~\% of peak
      intensity 0.9~K~\kms). The dashed contours are \CFINE \ of
      Stacey et al. (1993, contours 50, 70, 90~\% of peak intensity
      $3.9\cdot 10^{-3}$ erg cm$^{-2}$s$^{-1}$sr$^{-1}$).  The star
      symbols mark the positions of the Trapezium stars (\THEC \ 
      filled) and $\Theta _2$A Ori. Position used in Table 2 are
      marked as white circles and the phase center is marked by a
      cross.  }
    \label{fig1}
 \end{figure}
      
\begin{figure}[p]
  \hbox to \textwidth{
  \hfill
    \vbox{
    \psfig{figure=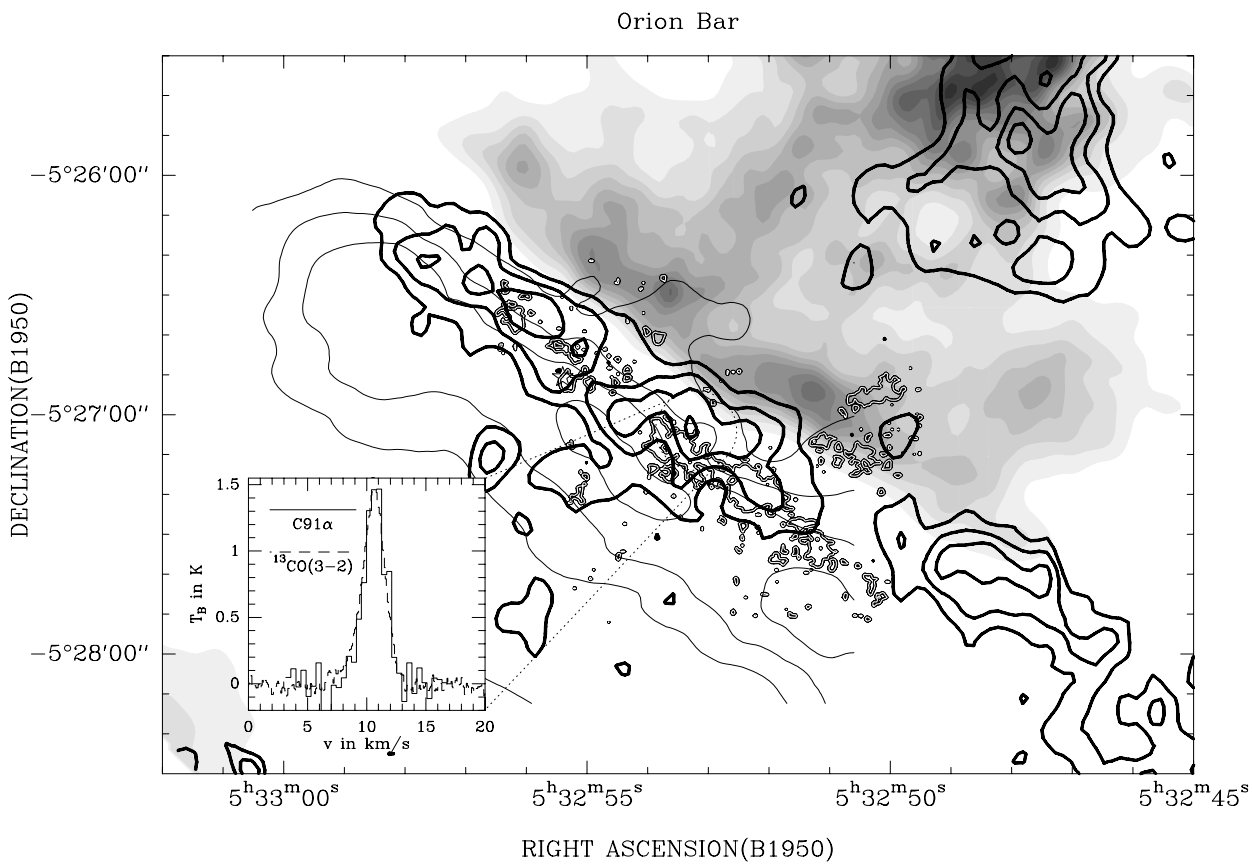,bbllx=10pt,bblly=20pt,bburx=580pt,bbury=860pt,width=18cm,angle=-90} 
         }
  \hfill
  }
    \caption
    {VLA 3.5 cm continuum (grey--scale, compared to \CNOA\ (VLA)
      integrated intensity (thick full contours: 30, 50, 70, 90~\% of
      peak intensity 5.5~K~\kms) and \THCO (3--2) integrated intensity
      from Lis \& Schilke (1997, thin contours, only the region
      around the Bar was mapped). The black and white contours are
      \MOLH (1-0 S(1)) image of van der Werf et al.  (1996). The
      insert in the lower left shows a comparison of \CNOA\ (smoothed
      to 20\arcsec) and \THCO\ (peak 29~K) spectra towards the
      indicated position in the Bar. }
    \label{fig2}
 \end{figure}
 
\begin{figure}[p]
  \hbox to \textwidth{
  \hfill  
    \vbox{
    \psfig{figure=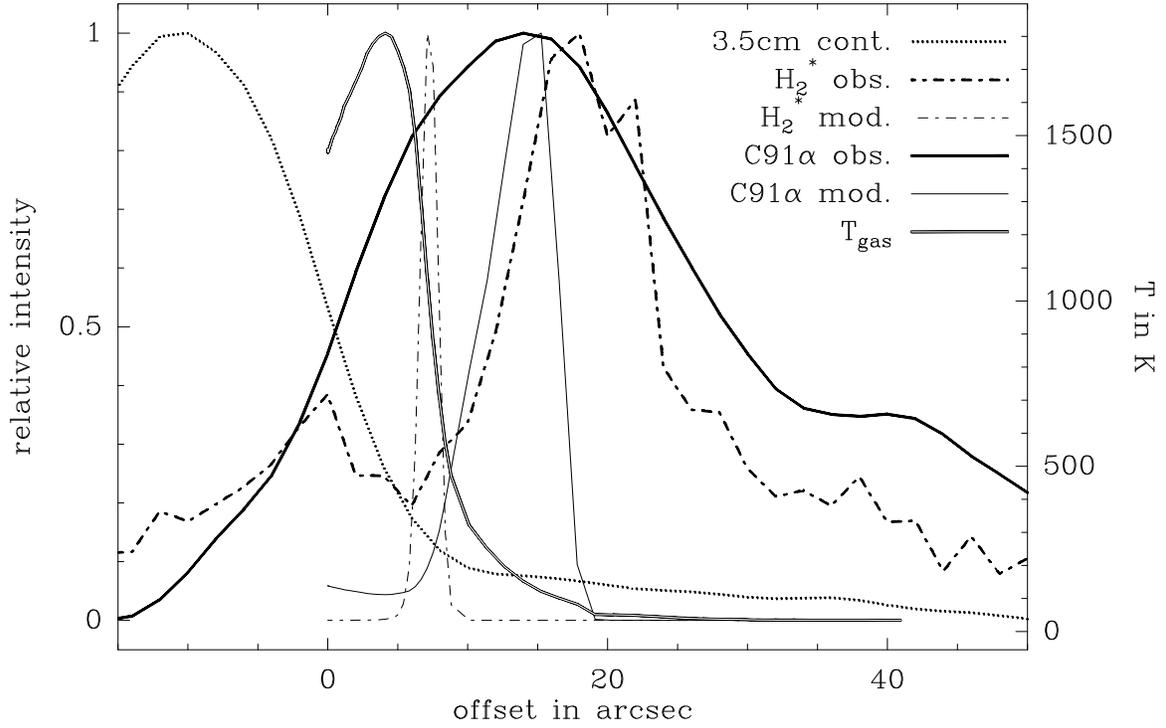,bbllx=0pt,bblly=0pt,bburx=580pt,bbury=830pt,width=16cm,angle=-90}
         }
  \hfill
  }
    \caption
    {Averaged intensity distribution of 3.5~cm continuum, \MOLH\ (1-0
      S(1)), and \CNOA\ across the Bar. The thin lines show for
      comparison results of PDR model calculations scaled to
      $n$(H)=$7\times 10^4$~\percc\ and $G_{0}=10^5$. The double line
      shows the model temperature distribution (right hand scale).}
    \label{fig1}
 \end{figure}
  
\newpage

\begin{table}[p]
\centering
\caption[vla parameters]
        {VLA Observing Parameters}
\vspace*{2mm}
\begin{tabular}{cc}
\hline
\hline
Rest Frequency (\CNOA ) &  8.589104~GHz   \\
Total Bandwidth         &  3.125 MHz  \\
Number of Channels      &  256    \\
Channel Separation      &  12 KHz (0.43~ km s$^{-1}$)  \\
Synthesized Beam FWHM   &   11.7 x 9.0~arcsec   \\
Primary Beam FWHM       &    5.2~arcmin \\
Phase Center of Map     &  $\alpha_{1950}$=05:32:51.30,
                           $\delta_{1950}$=-05:26:21.0   \\
Time On Source          &  7 hours    \\
\hline 
\end{tabular}
\label{line parameters}
\end{table}

\begin{table}[p]
\centering
\caption[intensities]
        {Integrated Intensities of \CNOA , \CSIA , and
        \CFINE \ at selected positions. The \CNOA\ data was smoothed
        to a 40\arcsec\ beam in order to match the different beam sizes.}
\vspace*{2mm}
\begin{tabular}{lcccc}
\hline
 Object & Offset  & I(\CNOA ) & I(\CSIA ) & I(\CFINE ) \\
 & arcsec  & K \kms & K \kms  & $10^{-4}$ erg cm$^{-2}$s$^{-1}$sr$^{-1}$ \\
\hline
Bar SW      & (-45, -85) & 1.7 & 0.75 & 22 \\
Bar center  & (45, -40)  & 2.1 & 0.79 & 28 \\
Orion South & (-55, 25)  & 2.6 & 0.83 & 36 \\
North of KL & (-30, 145) & 2.9 & 0.89 & 29 \\
\hline 
\end{tabular}
\end{table}
      
\end{document}